\documentclass[12pt]{article}
\usepackage{amssymb}

\def\espaciado{\lineskip .25 ex \baselineskip 4.8 ex \lineskiplimit 0 ex
               \parskip 2.5 ex plus .50 ex minus .25 ex}

\def\beq{\begin{equation}}
\def\eeq{\end{equation}}
\def\barr{\begin{eqnarray}}
\def\earr{\end{eqnarray}}

\newcommand{\ec}[1]{(\ref{#1})}
\def\CPn{\mathbb{C}P_n}

\begin{document}

\begin{titlepage}
\title{Extended Self-Dual Configurations as Stable Exact
Solutions in Born-Infeld Theory}

\author{\vspace{3mm}{J. Bellor\'{\i}n$^1$\thanks{{\tt jbellori@fis.usb.ve}}
\hspace{2mm} and \hspace{3mm} A. Restuccia$^2$\thanks{{\tt
arestu@usb.ve}}} \vspace{0.5mm}\\ {\fontsize{12}{14.4} ${}^1$
\textit{Dept. of Physics, Universidad de Oriente, Cuman\'a,}} \\
{\fontsize{12}{14.4} ${}^2$ \textit{Dept. of Physics, Universidad
Sim\'on Bol\'{\i}var, Caracas,}}
\\ {\fontsize{12}{14.4} \textit{Venezuela.}}}

\date{\normalsize July, 2000}
\maketitle

\begin{abstract}
\noindent A class of exact solutions to the Born-Infeld field
equations, over manifolds of any even dimension, is constructed.
They are an extension of the self-dual configurations. They are
local minima of the action for riemannian base manifolds and local
minima of the Hamiltonian for pseudo-riemannian ones. A general
explicit expression for the Born-Infeld determinant is obtained,
for any dimension of space-time.
\end{abstract} \thispagestyle{empty}
\end{titlepage}

\newpage
\setcounter{page}{2}
\espaciado

\section{Introduction}
The Born-Infeld action \cite{Born-Infeld, Gibbons+Ras., Zumino1,
Zumino2, Tseytlin} has been proposed as the effective theory of
open strings with Dirichlet boundary conditions \cite{Dai, Leigh,
Schim, Polchinski, Polchinski2}. The $D$-brane action arising in
this way is \beq S = \int d^{p+1}\xi
e^{-\phi}\sqrt{-\det\left[G_{\alpha\beta}+ B_{\alpha\beta}+
bF_{\alpha\beta}\right]} \label{DB action}\eeq where $G$, $B$ and
$\phi$ are the pullbacks of the 10 dimensional metric,
antisymmetric tensor and dilaton to the $D$-brane world-volume
while $F$ is the curvature of the world volume $U(1)$ gauge field
$A_\alpha$. In the full supersymmetric string theory this action
must be extended to a supersymmetric Born-Infeld (B-I) action. If
a number of assumptions are made the action \ec{DB action} becomes
the $U(1)$ Maxwell theory in $p+1$ dimensions which arises from
dimensional reduction of the $U(1)$ Maxwell theory in 10
dimensions with $N=1$ supersymmetry.

In order to study the highly nonlinear behavior of the theory
\ec{DB action} with respect to $F_{\alpha\beta}$, we will consider
\barr \phi &=& 0 \nonumber \\ B_{\alpha\beta}&=&0 \earr and
$G_{\alpha\beta}$ to be an external metric.

We first obtain a general explicit expression of the Born-Infeld
action valid for any dimension of the space-time.

We will then introduce a class of exact solutions to the field
equations of the B-I action. This class of configurations may be
considered an extension of the self-dual ones. We consider the B-I
action over a compact, riemannian manifold $M$ of even dimension
$D=2n$. We take all the products, with $m$ factors \beq P_m=
F\wedge\ldots\wedge F , \eeq where $F$ is the curvature of a
connection 1-form on a $U(1)$ principle bundle constructed over
$M$. We then define the extended self-dual configurations as the
connections 1-forms for which the set $\{P_m\hspace{1mm},
\hspace{1mm} m=0,1,\ldots,n\}$ is mapped into itself by the Hodge
dual operation. That is, they satisfy \beq ^*P_m \approx P_{n-m},
\label{dual*}\eeq for $m=0,\ldots,n$. Of course, the conditions in
\ec{dual*} are not all independent.

It is straightforward to see that the extended self-dual
configurations are solutions of Maxwell equations \beq d{^*F}=0.
\eeq

In fact, \beq ^*F={^*P}_1=kP_{n-1} \eeq and \beq dP_{n-1}=0. \eeq

We will show in this work that the extended self dual
configurations are also exact solutions of the Born-Infeld field
equations, moreover we will show they are strict minima of the
action. When the base manifold is pseudo-riemannian of the form
$M\times\mathbb{R}$, $M$ being compact riemannian, we will show
that the extended self-dual configurations over $M$ are the minima
of the corresponding Hamiltonian. As in the case of instanton
solutions to the self-dual equations, given a manifold with a
metric over it the solutions may or may not exist. We give in
section 3 examples where the extended self-dual configurations do
exist. The canonical connection 1-forms, introduced by Trautmant
in \cite{Trautmant}, over the Hopf fibring
\[ S_{2n+1} \longrightarrow \CPn \] are in particular examples of extended
self-dual configurations. It was also proven there that these
connections are solutions to the Maxwell equations.

\section{General formula for the determinant}
In this section we obtain the general formula for the determinant
of $g_{ab}+F_{ab}$ which we will extensively use in this work.

We may reexpress $g+F$, using the properties \barr g^{T} &=& g,
\nonumber \\ F^{T}&=&-F, \earr in the following way \beq
g+F=M^{T}DM+F, \eeq where $D$ is a diagonal matrix, while $M$ is
an orthogonal one.

We then have \beq \det\left(g+F\right)=\det\left(D+MFM^{T}\right)
\eeq where $MFM^{T}$ is again an antisymmetric matrix. We denote
the elements of $D$ as $\lambda_{a}\delta_{ab}$ and \[ G\equiv
MFM^{T}. \]

For a general $N\times N$ matrix, we have \barr
\det\left(D+G\right) &=& \sum\limits_{m}
\frac{1}{\left(2m\right)!\left(N-2m\right)!} \epsilon^{a_{1}\cdots
a_{m}c_{1}\cdots c_{m}e_{2m+1}\cdots e_{N}} \epsilon^{b_{1}\cdots
b_{m}d_{1}\cdots d_{m}e_{2m+1}\cdots e_{N}} \nonumber \\ & &\times
\lambda_{e_{2m+1}} \cdots \lambda_{N} G_{a_{1}b_{1}}\cdots
G_{a_{m}b_{m}}G_{c_{1}d_{1}}\cdots G_{c_{m}d_{m}} \label{deter1}
\earr

For a given element $\hat{E}$ of the set $E\equiv $\{ (
$e_{2m+1},\ldots, e_{N} ):$ each index may take values from 1 to
$N$ \}, the terms in \ec{deter1} \beq
\frac{1}{\left(2m\right)!}\epsilon^{a_{1}\cdots a_{m}c_{1} \cdots
c_{m}\hat{E}} \epsilon^{b_{1}\cdots b_{m}d_{1} \cdots
d_{m}\hat{E}}G_{a_{1}b_{1}}\cdots
G_{a_{m}b_{m}}G_{c_{1}d_{1}}\cdots G_{c_{m}d_{m}} \eeq represent
the determinant of a $2m\times 2m$ antisymmetric matrix. The
determinant of an antisymmetric $N\times N$ matrix is zero if $N$
is odd and for $N=2m$ one has the following expression, called the
\textit{pfaffian}, \beq
 \frac{1}{2^{2m}\left(m!\right)^2} \left(G_{a_{1}b_{1}}\cdots
G_{a_{m}b_{m}}\epsilon^{a_{1}\cdots a_{m}b_{1} \cdots
b_{m}\hat{E}} \right)^{2}. \eeq

We then have for \ec{deter1} the expression \barr
\det\left(D+G\right)&=&\sum_m\frac{1}{\left(N-2m\right)!
2^{2m}\left(m!\right)^{2}} \left(G_{a_{1}b_{1}}\cdots
G_{a_{m}b_{m}}\epsilon^{a_{1}b_{1}\cdots a_{m}b_{m}e_{2m+1}\cdots
e_{N}} \right)^{2} \nonumber \\ & &\times \lambda_{e_{2m+1}}\cdots
\lambda_{e_{N}}. \label{deter2} \earr

We may now \ec{deter2} in terms of the original $g$ and $F$
matrices, using that $\det M=1$, we obtain \barr
\det\left(g+F\right) &=&
\sum\limits_{m}\frac{1}{\left(N-2m\right)!2^{2m}\left(m!\right)^{2}}
\left(F_{a_{1}b_{1}}\cdots
F_{a_{m}b_{m}}\epsilon^{a_{1}b_{1}\cdots a_{m}b_{m}e_{2m+1}\cdots
e_{N}} \right) \nonumber \\ & & \times\left(F_{c_{1}d_{1}}\cdots
F_{c_{m}d_{m}}\epsilon^{c_{1}d_{1}\cdots c_{m}d_{m}l_{2m+1}\cdots
l_{N}} \right)g_{e_{2m+1}l_{2m+1}}\cdots g_{e_{N}l_{N}}. \earr

We now introduce \barr F &\equiv& F_{ab}dx^{a}\wedge dx^{b}
\nonumber\\ P_{m} &\equiv&\underbrace{F\wedge \ldots \wedge F}_{m}
= F_{a_{1}b_{1}}\cdots F_{a_{m}b_{m}}dx^{a_{1}} \wedge dx^{b_{1}}
\wedge \ldots \wedge dx^{b_{m}} \nonumber \\ {^*P_{m}} &=&
\sqrt{|g|}\epsilon_{a_{1}b_{1}\cdots a_{m}b_{m}e_{2m+1}\cdots
e_{N}} F^{a_{1}b_{1}}\cdots F^{a_{m}b_{m}}dx^{e_{2m+1}}\wedge
\cdots \wedge dx^{N}, \earr where $g\equiv\det g_{ab}$.

We then have, the final expression \beq
 \det\left(g+F\right) = |g|\sum
\limits_{m=0}^{n}\frac{1}{\left(N-2m\right)!2^{2m}\left(m!\right)^{2}}
{^*\left(P_{m} \wedge {^*P_{m}} \right)} \label{formula 1} \eeq
where $n=\frac{\left[N\right]}{2}$.

\section{A class of solutions for Born-Infeld field equations}
The Born-Infeld theory formulated over a Riemannian manifold $M$
may be described by the following $D$ dimensional action \beq
S\left(A\right)=\int\limits_{M}\left(\sqrt{\det\left(g_{ab}+bF_{ab}\right)}
-\sqrt{g}\right)d^{D}x \label{Accion} \eeq where $g_{ab}$ is an
external euclidean metric over the compact closed manifold $M$.
$F_{ab}$ are the components of the curvature of connection 1-form
$A$ over a $U(1)$ principle bundle on $M$.

We may express the $\det\left(g_{ab}+bF_{ab}\right)$, using the
general formula obtained in \ec{formula 1}, as \barr
\det\left(g_{ab}+bF_{ab}\right) &=& g\sum\limits_{m=0}^{n}
a_{m}b^{2m}{^*\left[P_{m}\wedge {^*P_{m}}\right]} \nonumber \\
&\equiv& gW. \earr

The first variation of \ec{Accion} is given by \beq \delta
S\left(A\right)=\int\limits_{M} W^{-\frac{1}{2}}\sum\limits_{m} m
a_{m} b^{2m} d\delta A\wedge P_{m-1}\wedge {^*P_{m}} \eeq which
yields the following field equations \beq \sum\limits_{m} m a_{m}
b^{2m} P_{m-1} \wedge d\left(W^{-\frac{1}{2}} {^*P_{m}}\right) =0.
\label{Ec. de Campo} \eeq

We introduce now a set $\mathcal{A}$ of $U(1)$ connection 1-forms
over $M$. They are defined by the following conditions, \beq
{^*P_{m}} \left(A\right) = k_{m} P_{n-m}\left(A\right)
\hspace{1cm} ,m=0,\ldots,n, \label{dual eq.} \eeq where
$n=\frac{D}{2}$, that is we assume the dimension $D$ of $M$ to be
an even natural number. \ec{dual eq.} may be interpreted as an
extension of the self duality condition. It is the condition that
the Hodge dual transformation maps the set
$\left\{P_{m}\hspace{1mm}, \hspace{3mm} m=0,\ldots,n\right\}$ into
itself.

We observe that these connections, if they exits in a $U(1)$
principle bundle over $M$, are solutions of the field equations
\ec{Ec. de Campo}.

In fact, \ec{dual eq.} implies \beq {^*\left[P_{m}\wedge
{^*P}_{m}\right]}=k_{m}{^*\left[P_{m}\wedge
P_{n-m}\right]}=k_{m}{^*P}_{n} \eeq but from \ec{dual eq.}, for
$m=n$, we obtain \beq {^*P}_{n}=k_{n} \eeq which is constant. We
thus have, for these connections, \beq W=\mathrm{constant}. \eeq

Finally, it results \beq
d\left(W^{-\frac{1}{2}}{^*P_{m}}\right)=k_{m}
W^{-\frac{1}{2}}d\left(P_{n-m}\right)=0, \eeq showing that
\ec{dual eq.}, if they exits, define a set of solutions to the
Born-Infeld field equations.

Let us analyze a particular case of \ec{dual eq.}. Let us consider
$n=\frac{D}{2}=1$. We then have \beq {^*P_{1}}={^*F} = k_{1}.
\label{mono1} \eeq

This solution represents a monopole connection over the $D=2$
manifold $M$. When $M$ is the sphere $S_{2}$, \ec{mono1} defines
the $U(1)$ connection describing the Dirac monopole on the Hopf
fibring $S_{3}\rightarrow S_{2}$. The constant $k_{1}$ is
determined from the condition \beq \int\limits_{M}
F=2\pi\times\mathrm{integer}\eeq which is a necessary condition to
be satisfied for a $U(1)$ connection, $F$ being its curvature.

This solution was extended to $U(1)$ connections over Riemann
surfaces of any genus in \cite{Mar+Res, Mar+Ovall+Res,
Mar+Res+Torr} where it was shown that they describe the minima of
the hamiltonian of the double compactified $D=11$ supermembrane
dual.

In general, if we consider $M$ to be $\CPn$ and the $U(1)$ Hopf
fibring \beq S_{2n+1}\rightarrow \CPn, \eeq there is a canonical
way to obtain $g_{ab}$ and the connection 1-form $A$ satisfying
\ec{dual eq.} from the metric over $S_{2n+1}$ \cite{Trautmant}. If
we consider local complex coordinates over $S_{2n+1}$,
$z_{\alpha}$ $\left(\alpha=0,1,\ldots,n\right)$, satisfying \beq
\bar{z}_{\alpha}z_{\alpha}=1, \eeq its line element
\[ d\bar{z}_{\alpha}d z_{\alpha} \]
may be decomposed into the line element of the base manifold
$\CPn$ and the canonical 1-form connection $\omega$, in the
following way \beq d\bar{z}_{\alpha}d z_{\alpha}=d
s^{2}-\omega^{2}. \eeq

In the coordinate system introduced in \cite{Trautmant} $\omega$
has the expression \[ \omega = u^{-1}d u + \frac{1}{2} \rho^{2}
\left(\bar{\zeta}_{a} d\zeta_{a} - \zeta_{a} d\bar{\zeta}_{a}
\right), \hspace{8mm} a=1,\ldots,n\] while the k\"ahlerian metric
is given by \[ ds^{2}= h_{ab} d\bar{\zeta}_{a} d\zeta_{b}
\] and the curvature \beq \Omega = \mathrm{i}d\omega = \mathrm{i}h_{ab}
d\bar{\zeta}_{a} \wedge d\zeta_{b} \label{kahler}\eeq with
\[ h_{ab} = \rho^{4} \left[ \delta_{ab} \rho^{-2}
-\zeta_{a}\bar{\zeta}_{b} \right] \] where
\[ \rho^{2} = \frac{1}{1+\bar{\zeta}_{a}\zeta_{a}}. \]

$u$ is the coordinate in the fiber over $\CPn$ and $\zeta_{a}$ the
local coordinates over $\CPn$.

It can be show that the K\"ahler 2-form \ec{kahler} with the
k\"ahlerian metric satisfies the extended self-dual conditions
\ec{dual eq.} \cite{Bellorin}.

\section{Minima of the Born-Infeld action}
We show in this section some properties of the set $\mathcal{A}$
of connection 1-forms we have introduced in the previous section.
On a given $U(1)$ principle bundle over $M$ there is at most one
$\hat{A} \in \mathcal{A}$ modulo flat connections on a trivial
bundle. In fact, we may consider the functional \beq
w\left(A\right)\equiv \int\limits_{M} \sum\limits_{m=1}^{n} a_{m}
P_{m}\wedge{^*P_{m}}. \eeq

We then have for $\hat{A}\in\mathcal{A}$ and any $A$ on the same
$U(1)$ principle bundle \beq w\left(A\right) =
w\left(\hat{A}\right)+ \int\limits_{M} \sum\limits_{m=1}^{n}
a_{m}\left[ P_{m} \left(A\right) -
P_{m}\left(\hat{A}\right)\right] \wedge{^*\left[
P_{m}\left(A\right)-P_{m}\left(\hat{A}\right)\right]} \label{D.2}
\eeq since \beq \int\limits_{M} \sum\limits_{m=1}^{n} a_{m}
P_{m}\left(A\right)\wedge{^*P_{m}\left(\hat{A}\right)}=\int\limits_{M}
\sum\limits_{m=1}^{n} a_{m} P_{m}
\left(\hat{A}\right)\wedge{^*P_{m}\left(\hat{A}\right)} \eeq
\ec{D.2} implies \beq w\left(A\right)\geq w\left(\hat{A}\right),
\eeq and \beq w\left(A\right) = w\left(\hat{A}\right) \eeq if and
only if \beq F\left(A-\hat{A}\right) = 0, \eeq where
$\left(A-\hat{A}\right)$ is a $U(1)$ connection on a trivial
bundle.

Moreover if $\hat{A}_{1}$ and $\hat{A}_{2}\in\mathcal{A}$ and are
on the same $U(1)$ principle bundle, then \beq
w\left(\hat{A}_{1}\right) \geq w\left(\hat{A}_{2}\right) \geq
w\left(\hat{A}_{1}\right). \eeq

Consequently, $\hat{A}_{2}-\hat{A}_{1}$ is a flat connection on a
trivial bundle.

We will denote $\{A\}$ the equivalence class of connections on a
given $U(1)$ principle bundle defined by elements which differ on
a flat connection on a trivial bundle.

The functional \ec{Accion} may be defined over the equivalence
classes $\{A\}$, assuming we are only considering $U(1)$
connections on the same principle bundle over $M$.

If there is $\hat{A} \in \mathcal{A}$ on a given $U(1)$ principle
bundle over $M$, $\{A\}$ is a strict local minimum of the
Born-Infeld action \ec{Accion}.

In fact, we obtain after some calculations \barr
S\left(\hat{A}+\delta A\right)-S\left(\hat{A}\right)&=&
\hat{W}^{-\frac{3}{2}}\int\limits_{M}\left[\hat{W}\sum\limits_{m=1}^{n}
a_{m} \delta P_{m}\wedge{^*\delta P_{m}} \right. \nonumber \\ & &
-\left. \sum\limits_{m=1}^{n} a_{m} \delta
P_{m}\wedge{^*\hat{P_{m}}}\sum\limits_{l=1}^{n}
a_{l}{^*\left(\delta P_{l}\wedge{^*\hat{P_{l}}}\right)}\right]
\nonumber \\ &=& \hat{W}^{-\frac{3}{2}}
\int\limits_{M}\sum\limits_{m=1}^{n} a_{m} \delta
P_{m}\wedge{^*\delta P_{m}} \nonumber \\ & & +
\hat{W}^{-\frac{3}{2}}\int\limits_{M}\left[\sum\limits_{m=1}^{n}
a_{m}\hat{P_{m}}\wedge{^*\hat{P_{m}}}\sum\limits_{l=1}^{n}
a_{l}{^*\left(\delta P_{l}\wedge{^*\delta P_{l}}\right)} \right.
\nonumber \\ & & -\left. \sum\limits_{m=1}^{n} a_{m} \delta
P_{m}\wedge{^*\hat{P_{m}}}\sum\limits_{l=1}^{n}
a_{l}{^*\left(\delta P_{l}\wedge{^*\delta P_{l}}\right)}\right].
\label{D.8} \earr

The second integral term in the right hand member of \ec{D.8} is
the discriminant of the roots of the second order equation in
$\lambda$ \beq \int\limits_{M}\left(P_{m}+\lambda \delta
P_{m}\right) \wedge {^*\left(P_{m}+\lambda \delta P_{m}\right)}=0.
\eeq

It is then greater or equal to zero.

Consequently we obtain \beq S\left(\hat{A}+\delta
A\right)-S\left(\hat{A}\right) \geq
\hat{W}^{-\frac{3}{2}}\int\limits_{M}\sum\limits_{m=1}^{n} a_{m}
\delta P_{m}\wedge{^*\delta P_{m}} \geq 0. \eeq

Moreover \[ S\left(\hat{A}+\delta A\right)=S\left(\hat{A}\right)
\] if and only if \[ F\left(A\right)=0, \] showing that $\{A\}$ is a strict
local minimum of \ec{Accion}.

\section{Minima of the Hamiltonian of Born-Infeld actions}
In this section we will show that the previous construction of the
minima of the Born-Infeld (B-I) actions over even $D$ dimensional
riemannian space may be performed for the Hamiltonian of a B-I
actions in $D+1$ pseudo-riemannian space-time. We will thus show
that the generalized self-dual configurations are the minima of
the Hamiltonian. We discuss this problem for a B-I action over
$M_2\times\mathbb{R}$ and $M_4\times\mathbb{R}$ space-times where
$M_2$ and $M_4$ are 2 and 4 dimensional compact riemannian spaces
respectively. The Hamiltonian analysis of the Born-Infeld theory
has been considered in several references \cite{Beng, Kh, Alex}.

We will consider the metrics in the ADM parametrization where
\barr g_{00} &=& -N^2+\gamma_{ab}N^aN^b \nonumber \\
g_{ab}&=&\gamma_{ab} \nonumber
\\ g_{0a} &=& g_{a0}=\gamma_{ab}N^b. \earr

It is well know that in all $p$-brane formulations in the light
cone gauge one obtains \barr N&=&1 \nonumber \\ N^a&=&0 .\nonumber
\earr

We will then work on that assumption which is in fact valid if we
think that the B-I action will be used as a model for $D$-brane
theories.

We start with the B-I action \beq S= \frac{1}{b^2}
\int\limits_{M_2\times\mathbb{R}} \left[\sqrt{-g} -
\sqrt{-\det\left(g_{\alpha\beta}+bF_{\alpha\beta}\right)}\right]
d^3x \label{accion*}\eeq

We thus obtain for the conjugate momenta to $(A_a)$, over
$M_2\times\mathbb{R}$, \beq \pi^a=
-\frac{\sqrt{\gamma}F^{0a}}{\left(1+\frac{1}{2}b^2F_{\alpha\beta}
F^{\alpha\beta}\right)^{\frac{1}{2}}} \label{pi2}\eeq where the
greek indices denote $\alpha=\left(0,a\right)$.

\ec{pi2} may be solved for the time derivatives, it yields \beq
\partial_0 A_a =\partial_aA_0+\frac{\left(1+\frac{1}{2}b^2F_{cd}
F^{cd}\right)^{\frac{1}{2}}}{\left(1+b^2\frac{1}{\gamma}\pi_c\pi^c\right)^
\frac{1}{2}} \frac{\pi_a}{\sqrt{\gamma}}. \label{despeje}\eeq

We finally obtain the Hamiltonian for the B-I action over
$M_2\times\mathbb{R}$ \beq H=\frac{1}{b^2}\int\limits_{M_2}
\left\{\left[1+\frac{1}{2}b^2F_{ab} F^{ab}\right]^\frac{1}{2}
\left[1+b^2\frac{\pi_c\pi^c}{\gamma}\right]^ \frac{1}{2}-1\right\}
\sqrt{\gamma}d^2x \label{hamilton2}\eeq subject to the first class
constraint \beq \phi\equiv \partial_a\pi^a=0 . \label{vinculo}\eeq

We will now analyze the minima of \ec{hamilton2}, \ec{vinculo}.
They are static solutions to the canonical field equations.

Variations of \ec{hamilton2} subject to \ec{vinculo} with respect
to $\pi_a$ yield the right hand side member of \ec{despeje}, $A_0$
being the Lagrange multiplier associated to \ec{vinculo}. A
solution to the stationary point condition is then \barr A_0 &=&
\mathrm{constant} \\ \pi^a&=&0. \label{picero}\earr

Variations with respect to $A_a$ give, the stationary point
condition \beq
\partial_a\left[ \frac{\left(1+b^2\frac{1}{\gamma}\pi_c\pi^c\right)^
\frac{1}{2}}{\left(1+\frac{1}{2}b^2F_{ab}
F^{ab}\right)^\frac{1}{2}} \sqrt{\gamma}
F^{ab}\right]=0.\label{ecHam}\eeq

If we now impose \ec{picero} into \ec{ecHam} we obtain the
Lagrangean field equations for the euclidean B-I action over $M_2$
discussed in section 2. We then conclude that the generalized
self-dual configurations are stationary points of the hamiltonian
of the B-I action. We will now consider the second variation of
\ec{hamilton2} subject to \ec{ecHam} around those solutions to
prove that these configurations are indeed local minima of the
hamiltonian. The only contribution to the second variation of
\ec{hamilton2} containing variations of $\pi^a$ is of the form
$\delta\pi_a\delta\pi^a$, since $\pi^a=0$. Its coefficient is
manifestly positive definite. The other contributions comes from
the second variations of \beq \frac{1}{b^2}\int\limits_{M_2}
\left\{\left[1+\frac{1}{2}b^2F_{ab} F^{ab}\right]^\frac{1}{2}
-1\right\} \sqrt{\gamma}d^2x, \eeq but this is exactly the B-I
action over euclidean space considered in section 2 and 3, where
it was show that these configurations are minima of it. We thus
conclude that these configurations are minima of the Hamiltonian
of the B-I action over $M_2\times\mathbb{R}$.

We may now analyze the problem over $M_4\times\mathbb{R}$. The
conjugate momenta to $A_a$ is given by \beq \pi^a =
\frac{-\sqrt{\gamma}\left(F^{0a}+b^2F^{0c}{^*F}_{cb}{^*F^{ab}}\right)}{
\left[1+\frac{1}{2}b^2F_{\alpha\beta}F^{\alpha\beta}+ \frac{1}{16}
b^4F^{\mu\nu}{^*F}_{\mu\nu\gamma}{^*F}^{\gamma\sigma\lambda}
F_{\sigma\lambda}\right]^{\frac{1}{2}}} . \label{pi4}\eeq

We proceed now to construct the Hamiltonian. It becomes difficult
to invert the time derivative of $A_a$ in terms of $\pi^a$. To
avoid this step we consider the following density \beq \mathcal{H}
= \frac{1}{b^2} \left[\sqrt{-\det\left(g_{\alpha\beta}+
bL_{\alpha\beta}\right)} - \sqrt{-\det g_{\alpha\beta}}\right]
\label{hamilton4}\eeq where $L_{\alpha\beta}$ is an antisymmetric
tensor, with \barr L_{0a} = \textrm{i} \frac{\pi_a}{\sqrt{\gamma}}
\nonumber \\ L_{ab}={^*F}_{ab}. \earr

$\mathcal{H}$ will be the correct Hamiltonian density if after
replacing of \ec{pi4} into the canonical action \beq S_c =
\int\limits_{M_4\times\mathbb{R}} \left[F_{0a}\pi^a -
\mathcal{H}\right]d^5x \label{accan}\eeq we recover the B-I
action.

We will show that this is the case. To do so, consider the
relations \barr -\det\left(g_{\alpha\beta}+bL_{\alpha\beta}\right)
&=& \det\left(\gamma_{ab}+bF_{ab}\right) + b^2\left(\pi_a\pi^a +
b^2\pi_b\pi^c F^{bd}F_{cd}\right)
\\ -\det\left(g_{\alpha\beta} + bF_{\alpha\beta}\right) &=&
\det\left(\gamma_{ab}+bF_{ab}\right) + \gamma b^2\left(
F_{0a}F^{0a}+ b^2F_{0a}{^*F}^{ab}{^*F}_{cb}F^{0c}\right). \earr

After several calculations we obtain \beq
\left[-\det\left(g_{\alpha\beta} + bF_{\alpha\beta}\right)\right]
\left[-\det\left(g_{\alpha\beta} + bL_{\alpha\beta}\right)\right]
= \left[\det\left(\gamma_{ab} + bF_{ab}\right)\right]^2. \eeq

We then use this formula to replace $\mathcal{H}$ in \ec{accan},
we also replace $\pi^a$ by \ec{pi4} we then exactly obtain the B-I
action \ec{accion*} over $M_4\times\mathbb{R}$.

We will now show that the extended self-dual fields are minimal
configurations of the Hamiltonian \beq H=\int\limits_{M_4}
\mathcal{H} d^4x. \eeq

The stationary point conditions for $H$ yield \beq \partial_a A_0
=\frac{\pi_a+b^2\pi_bF^{cb}F_{ac}}
{\sqrt{-\det\left(g_{\alpha\beta} + bL_{\alpha\beta}\right)}} ,
\eeq from which we consider the solutions \barr A_0 &=&
\mathrm{constant} \nonumber \\ \pi^a &=& 0, \nonumber \earr and
\beq \partial_a\left\{
\frac{\gamma\left[F^{ab}+\frac{1}{4}b^2F^{cd}{^*F}_{cd}{^*F}^{ab}
- b^2\frac{1}{\gamma}\pi_c\pi^b F^{ca} +
b^2\frac{1}{\gamma}\pi_c\pi^a F^{cb}\right]}
{\sqrt{-\det\left(g_{\alpha\beta} + bL_{\alpha\beta}\right)}}
\right\}=0 \eeq which reduces, after elimination of $\pi^a$, to
the B-I fields equations over $M_4$ considered in section 2.
Consequently the extended self-dual configurations are solutions
to the stationary point condition for $H$.

We now consider the second variation of $H$. The terms containing
$\delta\pi^a$ necessarily are of the form \[ \left(
\frac{1}{\gamma}\delta\pi_a\delta\pi^a+b^2\frac{1}{\gamma}
\delta\pi_b\delta\pi^c F^{bd}F_{cd}\right) \] with a positive
coefficient. They are then strictly positive. The others
contributions come from the second variation of \[ \frac{1}{b^2}
\int\limits_{M_4} \left[
\sqrt{\det\left(\gamma_{ab}+bF_{ab}\right)}-\sqrt{\det
\gamma_{ab}}\right]d^4x \] which is exactly the B-I action over
euclidean space analyzed in section 2 and 3. We then conclude that
the extended self-dual configurations over $M_4$ are local strict
minima of the hamiltonian \ec{hamilton4} provided that 1-form
connections over the same principle bundle are considered.

\section{Conclusions}
We constructed a class of stable exact solutions to the non-linear
Born-Infeld field equations. They leave invariant the set of
polinomic terms in the curvature \beq P_m = F\wedge\ldots\wedge F,
\hspace{8mm} m=0,\ldots,n \eeq under the Hodge dual operation. In
this sense they are an extension to other dimensions of the
self-duality configurations. We proved they are local minima of
the Born-Infeld action for riemannian manifolds and minima of the
Hamiltonian for pseudo-riemannian ones. The construction is
obtained by explicit use of a general expression for the
Born-Infeld determinant derived in section 2.

The extended self-dual configurations are defined for even
dimensions only. As in the case of self-dual ones they may or may
not exist on a given manifold with a given metric. They exist for
any k\"ahlerian manifold. We will discuss elsewhere related
configurations which minimize the Born-Infeld action over odd
dimensional space-times. We expect the extended self-dual
configurations will correspond to minima of related $D$-brane
actions. They certainly do correspond for the 2$D$-brane in $D=9$
as was shown in \cite{Mar+Ovall+Res}.

The paper \cite{Gibbons} appeared after we had finished our work.
The $D=4$ case in that paper overlaps with our analysis, which is
valid for any even dimension.

\end{document}